\documentclass{iau}
\usepackage{graphicx}
 \usepackage{natbib}

\title[Solar Wind during the Maunder Minimum] %% give here short title %%
{The State of the Solar Wind, Magnetosphere, and Ionosphere During the Maunder Minimum}

\author[Riley, Lionello, Linker \& Owens]   %% give here short author list %%
{ Pete Riley$^1$
Roberto Lionello$^1$
Jon A. Linker$^1$
and Mathew J. Owens$^2$
}

\affiliation{
$^1$Predictive Science Inc., 9990 Mesa Rim Rd, Suite 170, San Diego, CA 92121, USA \\email: {\tt pete@predsci.com} 
$^2$Space and Atmospheric Electricity Group, Department of Meteorology, University of Reading, Earley Gate, PO Box 243, Reading RG6 6BB, UK
}

\pubyear{2018}
\volume{340}  %% insert here IAU Symposium No.
\jname{Long – Term Datasets for the Understanding of Solar and Stellar Magnetic Cycle}
\editors{**** \& ****, eds.}
\begin{document}

\maketitle

\begin{abstract}

Both direct observations and reconstructions from various datasets, suggest that conditions were radically different during the Maunder Minimum (MM) than during the space era. Using an MHD model, we develop a set of feasible solutions to infer the properties of the solar wind during this interval. Additionally, we use these results to drive a global magnetospheric model. Finally, using the 2008/2009 solar minimum as an upper limit for MM conditions, we use results from the International Reference Ionosphere (ILI) model to speculate on the state of the ionosphere. The results describe interplanetary, magnetospheric, and ionospheric conditions that were substantially different than today. For example: (1) the solar wind density and magnetic field strength were an order of magnitude lower; (2) the Earth's magnetopause and shock standoff distances were a factor of two larger; and (3) the maximum electron density in the ionosphere was substantially lower.

%% add here a maximum of 10 keywords, to be taken form the file <Keywords.txt>
\end{abstract}

\firstsection % if your document starts with a section,
              % remove some space above using this command.
              
\section{Introduction}

%The ``Maunder Minimum'' (MM) was period of time  roughly spanning 1645 to 1715 when the observed number of sunspots effectively disappeared \citep{eddy76a}. We can summarize the available observations as follows: (1) Sunspots effectively disappeared for long intervals during 70-year period; (2) Eclipse reports suggest absence of visible K-corona, but appearance of F-corona; (3) Substantially reduced observations of aurora; and (4) Cosmic ray intensities at Earth inferred to be substantially higher. 

%\begin{figure}[b]
%\begin{center}
% \includegraphics[width=3in]{figures/riley15b-fig1.png} 
%\caption{Evolution of solar-related paramteters from 1600 through 2012: (a) Yearly Sunspot Number \citep{lockwood14a}; (b) Number of aurora per year \citep{rethly63a}; (c) Beryllium-10 \citep{berggren09a}; and (d) Carbon-14 measurements \citep{reimer04a}. Adapted from \citet{riley15b}.} 
%\label{fig-MM-overview}
%\end{center}
%\end{figure}

In this study, we build upon previous analysis \citep{riley15b} to speculate on the properties of the solar wind at 1 AU during the ``Maunder Minimum'' (MM)  as well as presenting some inferences for the likely configuration of the magnetosphere. Finally, we make some remarks about ionospheric conditions.

\section{Methods}

\citet{riley15b} described a wide range of ``observations'' during the Maunder minimum, ranging from descriptions of eclipses during the period between 1650 and 1715 \citep{eddy76a} to Be-10 \citep{berggren09a} and C-14 measurements \citep{reimer04a}. These were then used to define a set of candidate photospheric magnetic field configurations (see Figure~2 in \citet{riley15b}). 

%\subsection{Models}
%
%\begin{figure}[b]
%\begin{center}
%\includegraphics[width=5in]{figures/fig-mhd-scenarios.pdf} 
%\caption{Comparison of possible configurations of the Sun's photospheric magnetic field during the Maunder Minimum period: (a) CR 2085; (b) parasitic polarity ($\pm
%30$G) plus large-scale dipole (3.3G); (c) Large-scale dipole only (3.3G); (d) parasitic polarity ($\pm 30$G) plus large-scale dipole (1G); (e) parasitic polarity only ($\pm 10$G); and (f) parasitic polarity ($\pm 3.3$G). Adapted from \citet{riley15b}.} 
%\label{fig-mhd-scenarios}
%\end{center}
%\end{figure}

Using these candidate boundary conditions, we developed a set of MHD solutions (by integrating the time-dependent resistive MHD equations forward in time) and compared them with the limited observations. We concluded that the most likely state of the corona, at least during the deepest portion of the MM, was produced by a photospheric field composed of entirely ephemeral regions, likely of lower strength than observed today. In this study, we extrapolate these coronal solutions out to 1 AU, as well as applying a 1-D code to provide independent support for these 3-D results. 
To test the possible effects of the MM on the Earth's magnetosphere, we ran the BATS-R-US model \citep{zeeuw04a} as implemented at NASA's Community Coordinated Modeling Center (CCMC). Additionally, to explore the effects in the ionosphere, we ran the International Reference Ionosphere (IRI) model \citep{bilitza08a}. 

%\begin{eqnarray}
%\mathbf{ \nabla}\times \mathbf{B}& = & \frac{4 \pi}{c} \mathbf{ J} \\
%\mathbf{ \nabla}\times \mathbf{E}& = & -\frac{1}{c}
%\frac{\partial \mathbf{B}}{\partial t} \\ 
%\mathbf{ E} + \frac{\mathbf{ v}\times\mathbf{ B}}{c}& = & \eta \mathbf{ J}
%\label{in-eq} \\
%%
%\frac{\partial \rho}{\partial t} +\mathbf{ \nabla \cdot}(\rho \mathbf{ v})
%&=&0 \label{rho-eq}
% \\
%  %
%\frac{1}{\gamma -1} \left (
%\frac{\partial T}{\partial t}+ \mathbf{v}\cdot \mathbf{\nabla} T \right)
%&=&
%-T\mathbf{\nabla} \cdot \mathbf{v} +\frac{m}{2 k \rho} S \\
%%
%\rho \left (\frac{\partial \mathbf{ v}}{\partial t}
%+ \mathbf{ v \cdot}\mathbf{ \nabla} \mathbf{ v} \right ) &=&
%\frac{1 }{c} \mathbf{J}\times \mathbf{B}- \mathbf{ \nabla } (p+p_w)
%+ \rho\mathbf{g}
%+ \mathbf{\nabla }\cdot (\nu \rho \mathbf{ \nabla v})
%\label{mo-eq} \\
%S&=&
%(- \mathbf{\nabla}\cdot\mathbf{q}
%-  n_en_p{Q(T)}+{H_{\mathrm{ch}}}) \label{en-S} 
%\end{eqnarray}

%\begin{eqnarray}
% H = H_\mathrm{QS}+H_\mathrm{AR}
% \label{eq-jon} \\
%H_\mathrm{QS} = H^0_{\mathrm{QS}}  f(r) 
% \frac{ B_t^2 }{B(|B_r|+B_r^c )}
% \label{eq-fr} \\
%H_\mathrm{AR} = H_{\mathrm{AR}}^0 g(B) \left (\frac{B}{B_0}\right )^{1.2}
%\label{eq-gB}
%\end{eqnarray}
%

\section{Results}

Using the solar MHD model solutions, we inferred that the basic plasma/magnetic field properties of the MM heliosphere at Earth would be as summarized in Table~\ref{t:1}, which also compares these values with typical space era solar minimum conditions. Based on this, we infer that the average speed of the solar wind was probably a factor of 1.7 times slower (240 km s$^{-1}$), the radial magnetic field was an order of magnitude lower (0.1 nT), and the  density was approximately 24 times smaller (0.21 cm$^{-3}$).

%\begin{figure}[b]
%\begin{center}
%\includegraphics[width=5in]{figures/fig-mhd-fl.pdf} 
%\caption{As Figure~\ref{fig-mhd-scenarios} but showing a selection of magnetic field lines drawn from a grid separated by $10^{\circ}$ in latitude and longitude. Adapted from \citet{riley15b}.} 
%\label{fig-mhd-fl}
%\end{center}
%\end{figure}

\begin{table}[]
\centering
\begin{tabular}{lccc}
\hline
Quantity  & MM at 20 R\_s & MM at 1 AU & Typical 1 AU values   \\ 
                  \hline

$v_r$ (km/s)      & 240           & 240        & 400   \\ 
$B_r (nT)$             & $10^{-4}$ G   & 0.09 nT    & 1    \\
$n_p$ (cm$^{-3}$) & 25            & 0.21       & 5      \\ 
\hline
\end{tabular}
\caption{Inferred properties of the MM Sun: speed ($v_r$), radial magnetic field ($B_r$), and number density ($n_p$) are compared with typical values at 1 AU.}
\label{t:1}
\end{table}

\citet{riley10b} investigated conditions in the solar wind during the 2008/2009 minimum using a 1-D model that included the superradial expansion of the coronal magnetic field as well as correlation analysis between various observed parameters. In particular, we found: (1)  from Ulysses high-latitude/high-speed measurements: $B_r^{SW} \propto n_p^{SW}$; 2) from Wilcox/Ulysses measurements, the photospheric magnetic field within large polar coronal holes, $B_{ch} \propto B_r^{SW}$; and (3) from hydrodynamic simulations: $n_p^{SW}  \propto H$. Additionally, it had been established that coronal heating, $H \propto B$ \citep{pevtsov03a}. Taken together, these suggest $n_p^{SW} \propto B_{ch}$, supporting the 3-D MHD results that a substantial drop in the photospheric magnetic field should result in a substantial drop in the number density of the solar wind at 1 AU. 

Before we interpret the magnetospheric MHD simulations driven by these values, it is worth considering analytic approximations. We can compute the magnetopause stand-off distance using the following approximation  ({\it Spreiter et al.} (1966)): 
\begin{equation}
r \approx \sqrt[6]{\frac{2 B_0^2}{\mu_0 \rho v^2}}
\end{equation}
where $B_0$ is the Earth's dipole field strength and $\rho$ and $v$ refer to the density and speed of the solar wind. Using the values from Table~\ref{t:1}, we estimate the MM stand-off distance in relation to today's value to be: 
\begin{equation}
r_{MM} \approx 2 \times r_{2018}
\end{equation}

We can also estimate the dawn-dusk electric field applied by the solar wind across the magnetosphere as: 
\begin{equation}
E_y = -v_{SW} \times B_z
\end{equation}
which, for the MM conditions we infer, implies: $E_y^{MM} \approx 0.006 \times E_y^{2018}$, or a mere $1\%$ of today's value. 

To estimate the bow-shock stand-off distance, we must first estimate the solar wind Magnetosonic Mach Number: 
\begin{equation}
M_{ms} = \frac{v_{SW}}{\sqrt{v_A^2 + C_s^2}}
\end{equation}
which implies $M_{ms}^{MM} \approx M_{2018}$.  The  shock stand-off distance ({\it Spreiter et al.} (1966)) is thus: $\frac{\Delta}{r} \sim 1.1 \frac{n_{1}}{n_2} \sim 0.28$
where we have used: $ \frac{n_{2}}{n_{1}} = \frac{(\gamma + 1) M_{1}}{(\gamma - 1) M_{1}^2 + 2}$. 
This is relative to the magnetopause stand-off, and, thus, proportionately, the bow-shock sits ahead of the magnetosphere by the same fraction as today's configuration. Thus, with a typical value of $r_{shock} \sim14.5 R_E$ today, the MM shock stand-off distance would be  $\sim 29 R_E$. 

\begin{figure}[t]
\begin{center}
\includegraphics[width=4.0in]{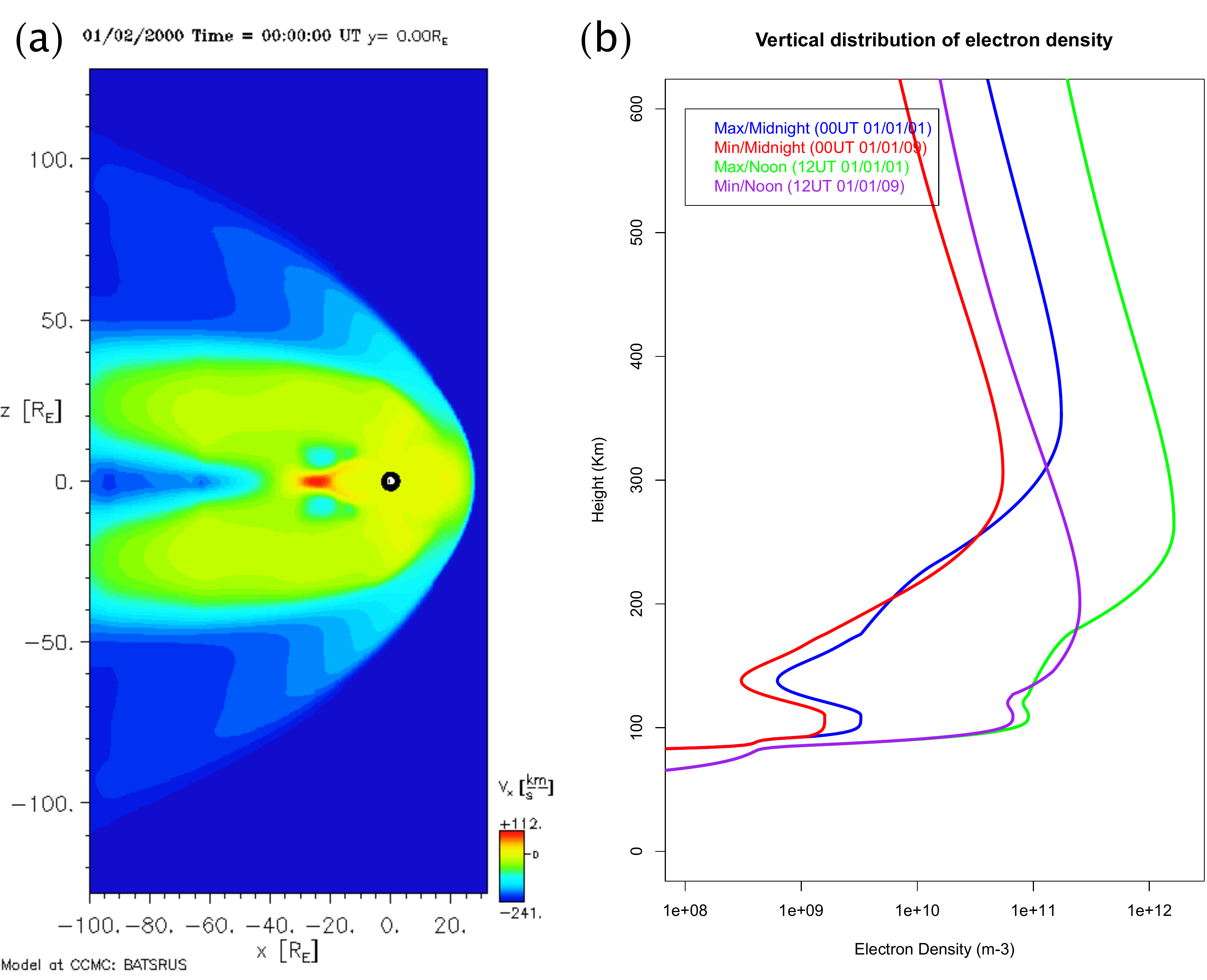} 
\caption{(a) BATS-R-US magnetospheric solution showing $V_x$ within the near-Earth environment. The Earth's location is marked by the black circle. (b) IRI ionospheric solution showing electron density as a function of height for solar minimum, maximum, noon, and midnight conditions.} 
\label{fig-iribats}
\end{center}
\end{figure}

To test these analytic results, we ran a global MHD simulation of the Earth's magnetosphere, setting the upstream solar values as in Table~\ref{t:1}, and allowing the simulation to reach equilibrium (Figure~\ref{fig-iribats}(a)). The magnetopause standoff distance is $29 R_E$, consistent with the analytic calculations. 

Finally, we ran the IRI to infer how the ionosphere may have appeared during the MM. Solutions for the 2008/2009 solar minimum (as an upper limit for MM conditions) as well as the 2001 solar maximum (at noon/midnight) are shown in Figure~\ref{fig-iribats}(b). These results suggest: (1) the maximum ionospheric density decreased significantly during MM; and (2) The F2 peak was located significantly lower. As pointed out by \citet{smithtro05a} over-the-horizon radio-wave propagation would have been restricted to notably lower frequencies, and the paths of radio waves would have been significantly modified. 

\section{Discussion}

%In this study, we have applied a range of numerical models to investigate the state of the solar corona, solar wind, magnetosphere, and ionosphere during the maunder minimum. We inferred that the most plausible state of the Maunder Minimum solar wind is as follows: (1) Low magnetic field strength; (2) Low solar wind speed; (3) Low proton density; (4) Modest variability; and (5) Very infrequent CME-related activity. 	

%Using these results, we drove a global MHD magnetospheric model to probe the likely state of the Earth's outer magnetosphere. We found that the magnetopause was approximately twice as far from the Earth as it is today ($29 R_E$ as oppose to $15 R_E$), and the dawn-dusk electric field was likely only a small fraction of what it is today ($\sim 1$\%). 
%On the other hand, the solar wind magnetosonic Mach number was approximately same as is is today. 

%Ionospheric conditions during the Maunder Minimum were driven primarily by differences in EUV flux that existed at the time, and not to changes in the plasma properties of the solar wind (at least during quiescent times, which dominated during this interval). Because of the reduced height of the peak electron density,  should such conditions have occurred during our technologically-dependent society, over-the-horizon radio-wave propagation would have been restricted to lower frequencies. On the other hand, because of the significantly reduced densities, GPS communications would have been greatly improved. 
%Finally, radio wave paths would have been substantially modified.  

The work summarized briefly here is only a starting point and suggests several potentially fruitful avenues to pursue in the future, to better understand the Maunder Minimum and the Earth's response to it. 
%First, our use of the IRI model allowed us only to compare the unique solar minimum of 2008/2009 with solar maximum conditions. However, as we have noted, conditions during the MM was likely substantially more weakened than this. Previous studies have attempted to infer the state of the ionosphere during the MM, but have relied on 1-D idealized calculations \citep{smithtro05a}. A more accurate and comprehensive study could now realistically use 3-D time-dependent models, such as the Thermosphere-Ionosphere-Exosphere-General-Circulation-Model (TIE-GCM) developed at NCAR to more accurately infer MM conditions in the thermosphere and ionosphere, driven by reasonable estimates for the solar EUV flux expected to have existed during this interval. In turn, it may be possible to infer what the solar spectrum may have looked like using the magnetograms we have inferred to have existed in this study as well as previous ones \citep{riley15b}.
For example, in the magnetosphere, our idealized simulations using the BATS-R-US code at the CCMC could be improved upon in several important ways. First, we assumed that the Earth's intrinsic magnetic field did not change appreciably. However, it is well known that 300-400 years ago, the dipole moment was  approximately 10\% larger than today \citep[e.g.][]{vogt07a}. This would modestly increase the disparity between the magnetospheric pressure and that of the  reduced solar wind, and, in particular, increase the stand-off distances inferred from this study. This, and other effects, however, would be relatively minor, and not change the inferences or conclusions reached here. Additionally, for simplicity, we ran the magnetospheric-only model. However, BATS-R-US has been coupled to several inner magnetospheric models, including the Rice Convection Model (RCM) and Comprehensive Ring Current Model (CRCM), as well as the Radiation Belt Environment (RBE) model. We chose not to include these because it was not clear that these models would be accurate under the new MM conditions imposed by the solar wind and global magnetospheric model. With the promising results presented here, however, the next logical step would be to add one of these components and explore the results.  
%Finally, this simulation tested the BATS-R-US code in a regime to which is rarely exposed (although some paleomagnetospheric calculations have investigated changes in the Earth's dipole moment \citep{vogt07a}). Thus, it would be instructive to run at least one other global MHD code (e.g., the LFM or OpenGGCM) to verify that that the results are not sensitive to any nuances in the algorithms. 

%Our ionospheric calculations were limited to the IRI model, which requires that the user input a valid date/time for which measurements exist. It is thus not possible to reduce EUV inputs to levels thought to have been present during the MM. We are currently running more sophisticated simulations using the Thermosphere Ionosphere Electrodynamics General Circulation Model (TIE-GCM) using a range of lower EUV values to investigate the spatial and temporal structure of the ionosphere during MM-like conditions. 

\section{Acknowledgements}
PR, RL, and JAL gratefully acknowledge support from NASA (Living with a Star program, NNX15AF39G) and NSF (FESD program, AGS-1135432). 
MO is funded by Science and Technology Facilities Council (STFC) grant ST/M000885/1. The magnetospheric simulation results have been provided by the Community Coordinated Modeling Center at Goddard Space Flight Center through their public Runs on Request system (http://ccmc.gsfc.nasa.gov). The BATS-R-US Model was developed by the Center for Space Environment Modeling group at the University of Michigan. The IRI model results were provided by the IRI-2012 code through the Virtual Ionosphere, Thermosphere, Mesosphere Observatory (VITMO). 

%\bibliographystyle{aipproc}
%\bibliography{/Users/pete/Dropbox/manuscripts/references/riley-refs-v3-1}  

\end{document}